\documentclass[journal]{IEEEtran}
\usepackage[utf8]{inputenc}
\usepackage[english]{babel}
\usepackage{epsfig}
\usepackage{amsfonts}
\usepackage{ifpdf}
\usepackage{multirow}
\usepackage{amssymb}
\usepackage{amsthm}
\usepackage{mathtools}
\usepackage{mathrsfs}
\usepackage{textgreek}
\usepackage{grffile}
\usepackage{xcolor}
\usepackage{bm}
\usepackage{cite}
\usepackage{setspace}

\usepackage[usestackEOL]{stackengine}

  \stackMath

\ifCLASSINFOpdf
\else
\fi
\usepackage{amsmath}
\usepackage{algorithmic}
\usepackage{algorithm}
\usepackage{graphicx}

\newcommand{\norm}[1]{\left\lVert#1\right\rVert}

\ifCLASSOPTIONcompsoc
  \usepackage[caption=false,font=normalsize,labelfont=sf,textfont=sf]{subfig}
\else
  \usepackage[caption=false,font=footnotesize]{subfig}
\fi

\usepackage{stfloats}
\begin{document}
%\title{Massive MIMO Enabled Non-Coherent Modulation With One-bit ADCs}

%\title{Joint Positioning and Communication with Reconfigurable Intelligent Surfaces and Time-Resistant   }

\title{Impact of Model Mismatch on DOA Estimation with MUSIC: Near-Field and Far-Field}
\author{Don-Roberts~Emenonye,
        Harpreet~S.~Dhillon,
        and~R.~Michael~Buehrer% <-this % stops a space
\thanks{D.-R. Emenonye, H. S. Dhillon, and R. M.  Buehrer are with Wireless@VT,  Bradley Department of Electrical and Computer Engineering, Virginia Tech,  Blacksburg,
VA, 24061, USA. Email: \{donroberts, hdhillon, rbuehrer\}@vt.edu. The support of the US National Science Foundation (Grants ECCS-2030215 and CNS-2107276) is gratefully acknowledged. 
}
}

\maketitle
\IEEEpeerreviewmaketitle
\begin{abstract}
There has been substantial work on developing variants of the multiple signal classification (MUSIC) algorithms that take advantage of the information present in the near-field propagation regime. However, it is not always easy to determine the correct propagation regime, which opens the possibility of incorrectly applying simpler algorithms (meant for far-field) in the near-field regime.  Inspired by this, we use simulation results to investigate the performance drop when there is a mismatch between the signal model in the MUSIC algorithm and the propagation regime. For direction of arrival (DOA) estimation, we consider the cases when the receiver is in the near-field region but uses i) the near-field model, ii) the approximate near-field model (ANM) model, and iii) the far-field model to design the beamforming matrix in the MUSIC algorithm. We also consider the case when the receiver is in the far-field region, and we use the correct far-field model to design the beamforming matrix in the MUSIC algorithm. One contribution is that in the near-field, we have quantified the loss in performance when the ANM and the far-field model are used to create the beamforming matrix for the MUSIC algorithm, causing a reduction in estimation accuracy compared to the case when the correct near-field model is used to design the beamforming matrix. Another result is that in the near-field, when we incorrectly assume that the receiver is in the far-field and subsequently use the far-field beamforming matrix, we underestimate the DOA estimation error. Finally, we show that the MUSIC algorithm can provide very accurate range estimates for distances less than the Fraunhofer distance. This estimate gradually becomes inaccurate as the distances exceed the Fraunhofer distance.
\end{abstract}

\section{Introduction}

The ubiquitous availability of multiple antenna receivers enables the estimation of the direction of arrival (DOA) of wireless signals  \cite{8736783,elzanaty2023near,661337}.
% in the presence \cite{emenonye2022fundamentals,emenonye2022ris,emenonye2023_ICC_conf,emenonye2023_ICC_conf_workshop} and absence of reconfigurable intelligent surfaces (RIS) \cite{8736783,elzanaty2023near,661337}.
  This DOA estimation of wireless signals is also usually performed assuming that the source and destination nodes operate in the far-field propagation regime where the signal wavefront is assumed to be planar and the signal amplitude is assumed to remain constant across the receive antennas. However, recent advances in wireless systems involve using higher frequency bands, which allow for anchors with a more significant number of antennas. This combination of higher frequency bands and a larger number of antennas results in the extension of the near-field region. This increased relevance of the near-field region has resulted in works that provide both bounds on the estimation accuracy for DOA and algorithms to achieve these bounds \cite{emenonye2022fundamentals,emenonye2022ris,emenonye2023_ICC_conf,emenonye2023_ICC_conf_workshop}. However, the question of how the nodes know the propagation regime has not been answered, and it is possible that the far-field model will be used while the nodes are actually in the near-field region. Hence, in this paper, we investigate the effect of model mismatch (using the far-field model while in the near-field) on the performance of the  multiple signal classification (MUSIC) algorithm. It is important to note that model mismatch between the near and far fields has been studied in the context of spectral efficiency \cite{10104065}.

\subsection{Prior Art}
Prior literature on DOA estimation in the near-field involves developing algorithms that take advantage of the extra information in the spherical wavefront while minimizing complexity. Variants of the MUSIC algorithm have been proposed for joint DOA and range estimation \cite{8359308,8723631,8419297,7579547,8641299,9159941}. The authors in \cite{8359308,8723631,8419297,7579547,8641299,9159941} focus on the uniform linear array, with authors in \cite{8359308,7579547} concentrating on reducing the complexity, \cite{8723631} focusing on estimating the polarization, \cite{8419297} focusing on estimation with unknown path gains, \cite{8641299} on tracking multiple near-field sources, and \cite{9159941} extends \cite{8641299} to account for unknown spatially non-uniform noise. Mixed near and far field sources are handled with MUSIC in \cite{6451125,5200332,9659805}. In \cite{6451125}, the correlation matrix is replaced with a cumulant matrix, and \cite{9659805} handles the problem when the sample size is on the same order as the number of receive antennas. The cumulant matrix, in addition to the estimation of signal parameters via rotational invariant techniques (ESPRIT), is used for near-field in \cite{661337}. Machine learning is used for estimating the DOA from mixed near and far field sources in \cite{9477127} and from only near-field sources in \cite{9136773}.

While the prior art has primarily focused on modifying the current MUSIC algorithm to handle the near-field while minimizing complexity, there has yet to be any research on the mismatch that occurs when we utilize the incorrect signal model matrix in the MUSIC algorithm. More specifically, there has been little research on the mismatch that occurs when using a signal model that differs from the actual propagation regime.  

\subsection{Contribution}
\label{Impact_of_Model_Mismatch_on_DOA_Estimation_with_MUSIC_contribution}
In this paper, we study the performance of the MUSIC algorithm for DOA and range estimation under different combinations of the signal models used in the MUSIC algorithm and the propagation regimes. We show the impact of mismatch when the far-field or ANM model is used in the MUSIC algorithm while the destination is in the near-field region. We demonstrate that the best DOA and range performance occurs when we use the near-field model in the MUSIC algorithm, and the destination is in the near-field region. For DOA estimation, we show that using the ANM in the MUSIC algorithm is not very accurate in the near-field but becomes accurate at distances greater than the Fraunhofer distance. 

We also establish that using the far-field model in the MUSIC algorithm while incorrectly assuming that the destination node is experiencing far-field propagation leads to a severe underestimation of the DOA estimation error. Interestingly, simulations indicate that the DOA estimation error when using the near-field model in the MUSIC algorithm while in the near-field converges to using the far-field model in the MUSIC algorithm while in the far-field propagation regime. Finally, we show that using the near-field model in MUSIC for range estimation is very accurate at distances smaller than the Fraunhofer distance. This error in range estimation increases drastically at distances more significant than the Fraunhofer distance.

\section{System Model}
We consider $N_K$ single antenna sources, where the $k^{\text{th}}$ source is located at $\bm{p}_{k} = [x_k, y_k, z_k]^{\mathrm{T}}$ where $k \in [1,2,3, \cdots, N_K]$, and the locations of the sources are defined with respect to the global origin. The destination node is located at $\bm{p}_{U} = [x_U, y_U, z_U]^{\mathrm{T}}$, with its $u^{\text{th}}$ antenna element located at $\bm{s}_{u} = [x_u, y_u, z_u]^{\mathrm{T}}$. The location of the centroid of the destination is defined with respect to the global origin, while the location specified by $\bm{s}_{u}$  is defined with respect to $\bm{p}_{U}$.  The location of the  $u^{\text{th}}$ antenna element on the destination node can be defined with respect to the global origin as $\bm{p}_u = \bm{p}_U +  \bm{s}_u$. The destination located at ${\bm{p}}_{{U}}$ can be defined with respect to the $k^{\text{th}}$ source located at ${\bm{p}}_{{k}}$ as ${\bm{p}}_{{U}} = {\bm{p}}_{{k}}     +d_{{k} {U}} \bm{\Delta}_{{k} {U}},$ where  $d_{{k} {U}}$ is the distance from point ${\bm{p}}_{{k}}$ to point ${\bm{p}}_{{U}}$, and $\bm{\Delta}_{{k} {U}}$ is the corresponding unit direction vector where $\bm{\Delta}_{{k} {U}} = [\cos \phi_{{k} {U}} \sin \theta_{{k} {U}}, \sin \phi_{{k} {U}} \sin \theta_{{k} {U}}, \cos \theta_{{k} {U}}]^{\mathrm{T}}$. The locally defined points on the destination array can be collected in matrix form as  $ {\bm{S}}_{U} = [{\bm{s}}_1, {\bm{s}}_2, \cdots, {\bm{s}}_{N_U} ].$

Each source communicates with the destination node through $T$ transmissions, such that the signal received at the destination node during the $t^{\text{th}}$ transmission is
\begin{equation}
\label{equ:receive_processing}
\begin{aligned}
\bm{y}_{t} &=  \bm{H}_{}^{}  \bm{x}_{t}
+ \bm{n}_{t}, \\ &=   \bm{\mu}^{}_{t}+  \bm{n}_{t}.
\end{aligned}
\end{equation}
In the above equation, $\bm{\mu}^{}_{t}$ is the noise-free part of the signal received during the $t^{\text{th}}$ transmission, and $\bm{n}_{t}\sim \mathcal{C}\mathcal{N}(0, N_0)$  represents the thermal noise local to the destination's antenna array. Here, $\bm{x}_{t} \in [{x}_{1,t},{x}_{2,t}, \cdots, {x}_{N_K,t}]^{\mathrm{T}}$ is the collection of the symbols transmitted from the $N_K$ sources during the $t^{\text{th}}$ transmission. The entries on the $u^{\text{th}}$ row and $k^{\text{th}}$ column are $[\bm{H}]_{[u,k]} = \beta_{u} e^{-j 2 \pi f_{c} \tau_{ ku}}$. In this channel entry, $\beta_{u}  = {\beta_{u}}_{\mathrm{R}} + j{\beta_{u}}_{\mathrm{I}}$ is the complex path gain, $f_c$ is the operating frequency, and $\tau_{ ku}$ is the propagation delay from the  $k^{\text{th}}$ source located at $\bm{p}_k$ to the receive antenna located at $\bm{p}_u$ on the destination's antenna array. The delay is defined as  $\tau_{ ku} = \frac{\norm{\bm{p}_{u} - \bm{p}_{k}}}{c}$, and this definition captures the spherical curvature wavefront present in the signal received at the destination. When substantial spherical curvature in the signal received at the destination is observable at the destination node, the destination is said to be experiencing the effects of near-field propagation. However, when substantial curvature can not be observed by the destination's antenna array, the destination is said to be experiencing the effects of far-field propagation. The effects of far-field propagation can be modeled through a plane wave approximation. In this case, the delay can be approximated as $\tau_{ku} = \tau_{kU} + \Delta_{kU}^{\mathrm{T}} ({\bm{s}}_{u}) / c.$ Whether, the destination experiences near or far field propagation is dependent on the operating frequency and antenna array size, as evident by the boundary specified by the Fraunhofer distance, $d_{\mathrm{f}}=2 D^2 / \lambda$. Here, $\lambda$ indicates the wavelength of the signal and  $D$  the maximum diameter among the source and destination surface diameters \cite{emenonye2022ris}. While (\ref{equ:receive_processing}) adequately describes the signal in the near-field, it is computationally more efficient to approximate the signal in the near-field with a parabolic wave, not a spherical wave. In this paper, this case will be termed the approximate near-field model (ANM). The delay in this case can be written as $\tau_{ ku} = \frac{\norm{\bm{p}_{u} - \bm{p}_{k}}}{c} = \frac{\bm{s}_{k} + d_{{k} {U}} \bm{\Delta}_{{k} {U}} }{c}$ and, the entries on the $u^{\text{th}}$ row and $k^{\text{th}}$ column of the channel matrix 
are $[\bm{H}]_{[u,k]}^{\text{}} = \beta_{u} e^{-j 2 \pi f_{c} \frac{\bm{s}_{k} + d_{{k} {U}} \bm{\Delta}_{{k} {U}} }{c}}$. 

With the far-field delay approximation, the received signal equation can be approximated by
\begin{equation}
\label{equ:far_field_receive_processing}
\begin{aligned}
\bm{y}_{t} &= \sum_{k = 1}^{N_K} \bm{a}_{Uk}(\Delta_{kU})   e^{-j 2 \pi f_{c}
\tau_{kU}} \bm{x}_{kt} + \bm{n}_{t}, \\
&= \bm{H}^{FF}   \bm{x}_{t} + \bm{n}_{t},
\end{aligned}
\end{equation}
where  $\bm{a}_{Uk}(\Delta_{kU}) = e^{-j 2 \frac{\pi}{\lambda} {\bm{S}}_{U}^{\mathrm{T}} \Delta_{kU} }$, $[\bm{a}_{Uk}(\Delta_{kU})]_{u} = \beta_u e^{-j 2 \frac{\pi}{\lambda} {\bm{s}}_{u}^{\mathrm{T}} \Delta_{ku}}$, and $[\bm{H}^{\text{FF}}]_{[u,k]} = \beta_u e^{-j 2 \frac{\pi}{\lambda} {\bm{s}}_{u}^{\mathrm{T}} \Delta_{ku}}e^{-j 2 \pi f_{c}
\tau_{kU}}$. As opposed to the near-field model and ANM, the delay in the far-field model is constant across all antennas, and hence cannot be estimated without more information.

%, \cite{6111312}

%The DOAs of Mixed sources are handled in \cite{}.

%In \cite{8932416}, the MUSIC algorithm is modified to improve estimation accuracy at low SNRs (signal-to-noise ratios). The estimation of signal parameters via rotational invariant techniques (ESPRIT) is extended to estimate polarization in the near-field \cite{9682707}. In \cite{8641299}, the ESPRIT algorithm is modified to obtain an algorithm for tracking sources in the near-field. The algorithm is modified in \cite{9159941} to account for the case of nonuniform additive noise.

%Other algorithms that are not simple variants of MUSIC and ESPRIT are found in \cite{}
%Advances in wireless systems usually involve using higher frequency bands, and these higher frequencies allow for anchors with a larger number of antennas. The combination of higher frequency bands and a larger number of antennas results in the extensions of the near-field region, and this increased relevance of the near-field region has resulted in recent works studying its effect in systems with \cite{emenonye2022fundamentals,emenonye2022ris,emenonye2023_ICC_conf,emenonye2023_ICC_conf_workshop} and without reconfigurable intelligent surfaces \cite{8736783,elzanaty2023near,661337}. 

%However, most of these works that incorporate

\begin{figure}[htb!]
\centering
%\subfloat[]{\includegraphics[height=1.8in, width= 1.5in]{Results/ber_ring_ratio_parallel.eps}
\subfloat[]{\includegraphics[width=\linewidth]{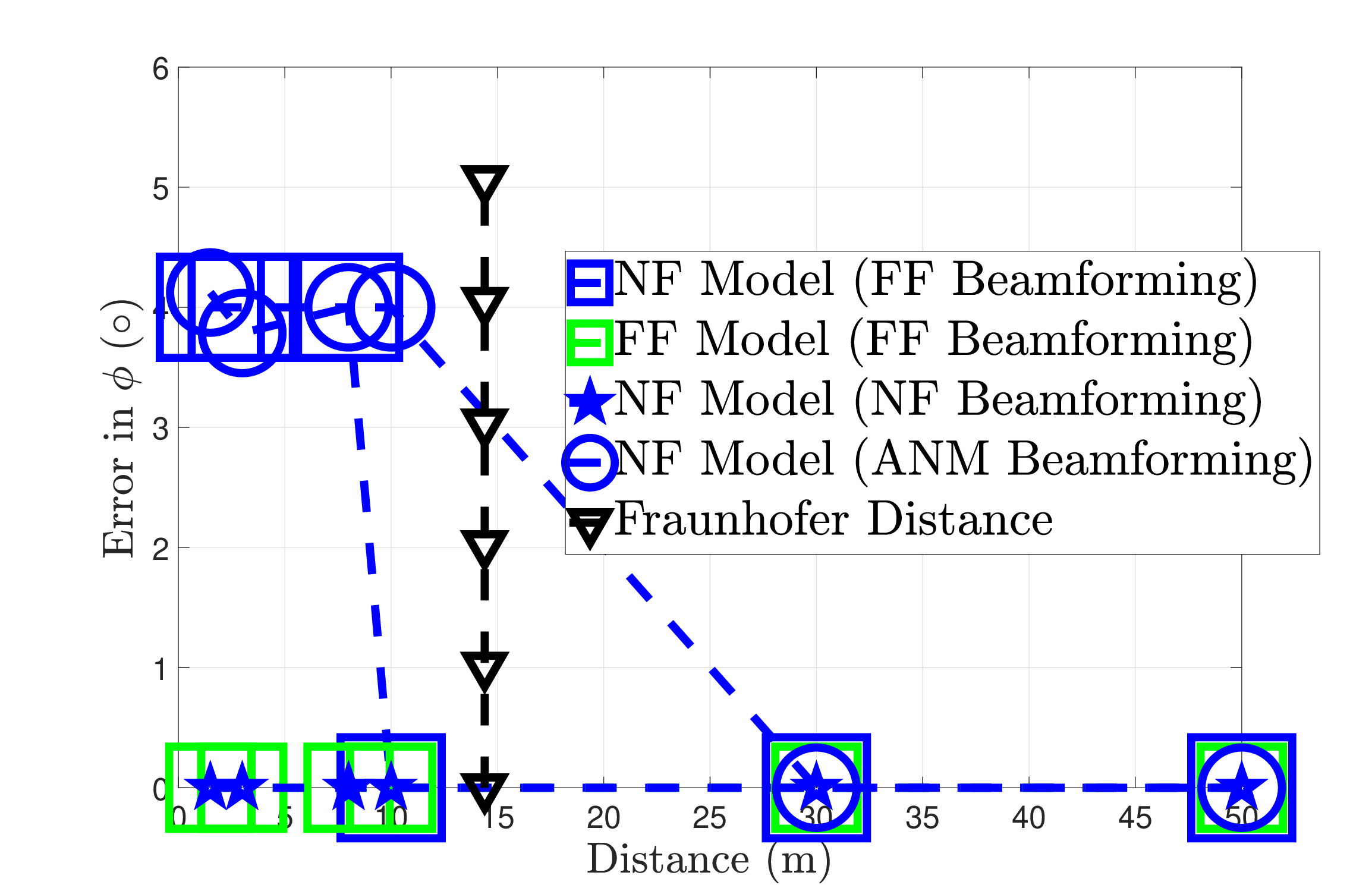}
\label{Figure/NB_2_NS_1280_Nangles_8_fc_1_SNRInd_4_NU__a/Phi_NB_2_NS_1280_Nangles_8_fc_1_SNRInd_4_NU_3.eps}}
\hfil
\subfloat[]{\includegraphics[width=\linewidth]{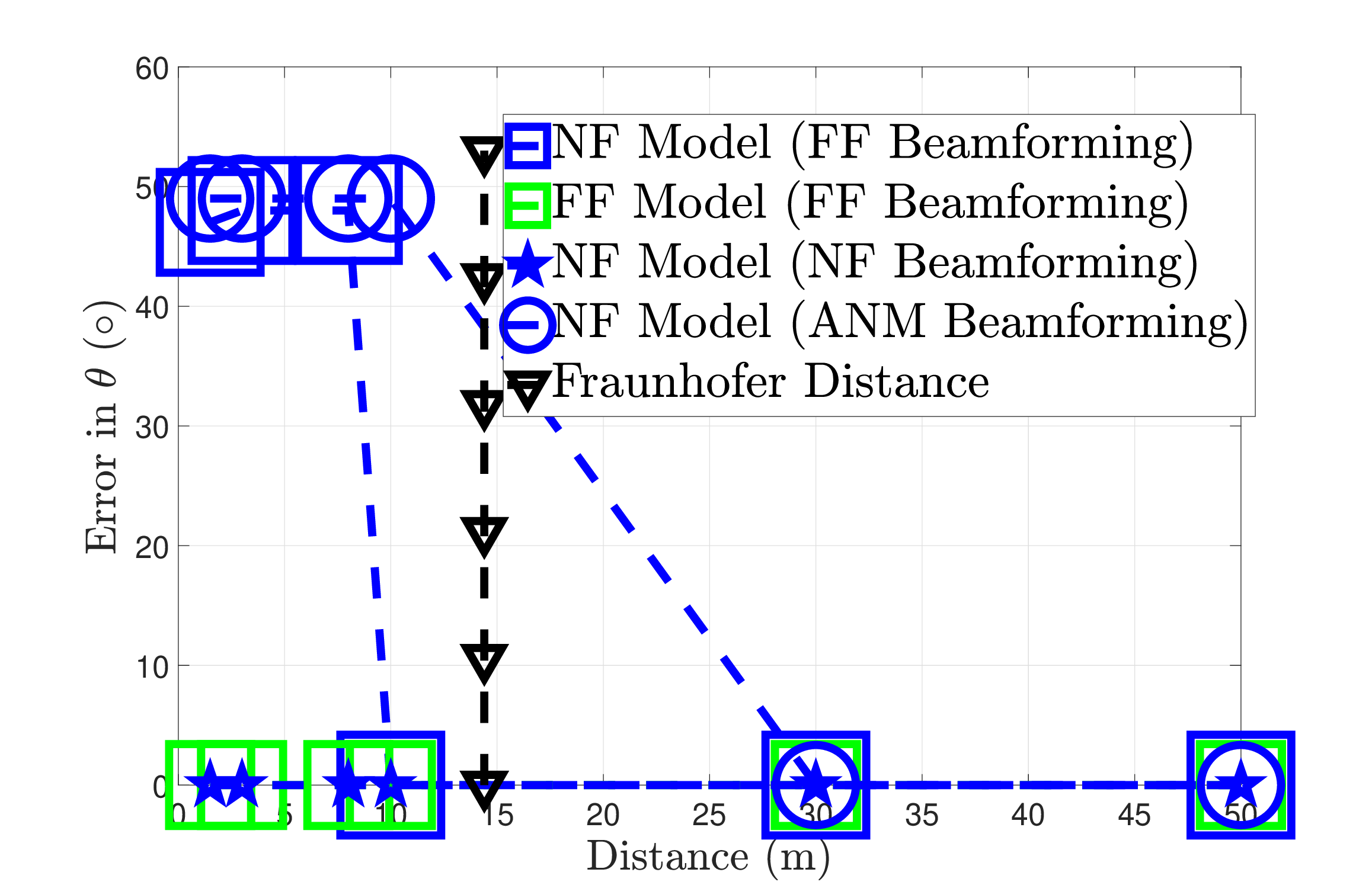}
\label{Figure/NB_2_NS_1280_Nangles_8_fc_1_SNRInd_4_NU__a/Theta_NB_2_NS_1280_Nangles_8_fc_1_SNRInd_4_NU_3.eps}}
\caption{Error in estimating azimuth and elevation angles with $f_c = 3 \times 10^9$, SNR of $30 \text{ dB}$, and $N_U = 144$. This figure highlights the different combinations of the received signal models and the various beamforming vectors.}
\label{Figure/NB_2_NS_1280_Nangles_8_fc_1_SNRInd_4_NU__a/Theta_Phi_NB_2_NS_1280_Nangles_8_fc_1_SNRInd_4_NU3}
\end{figure}

\begin{figure}[htb!]
\centering
%\subfloat[]{\includegraphics[height=1.8in, width= 1.5in]{Results/ber_ring_ratio_parallel.eps}
\subfloat[]{\includegraphics[width=\linewidth]{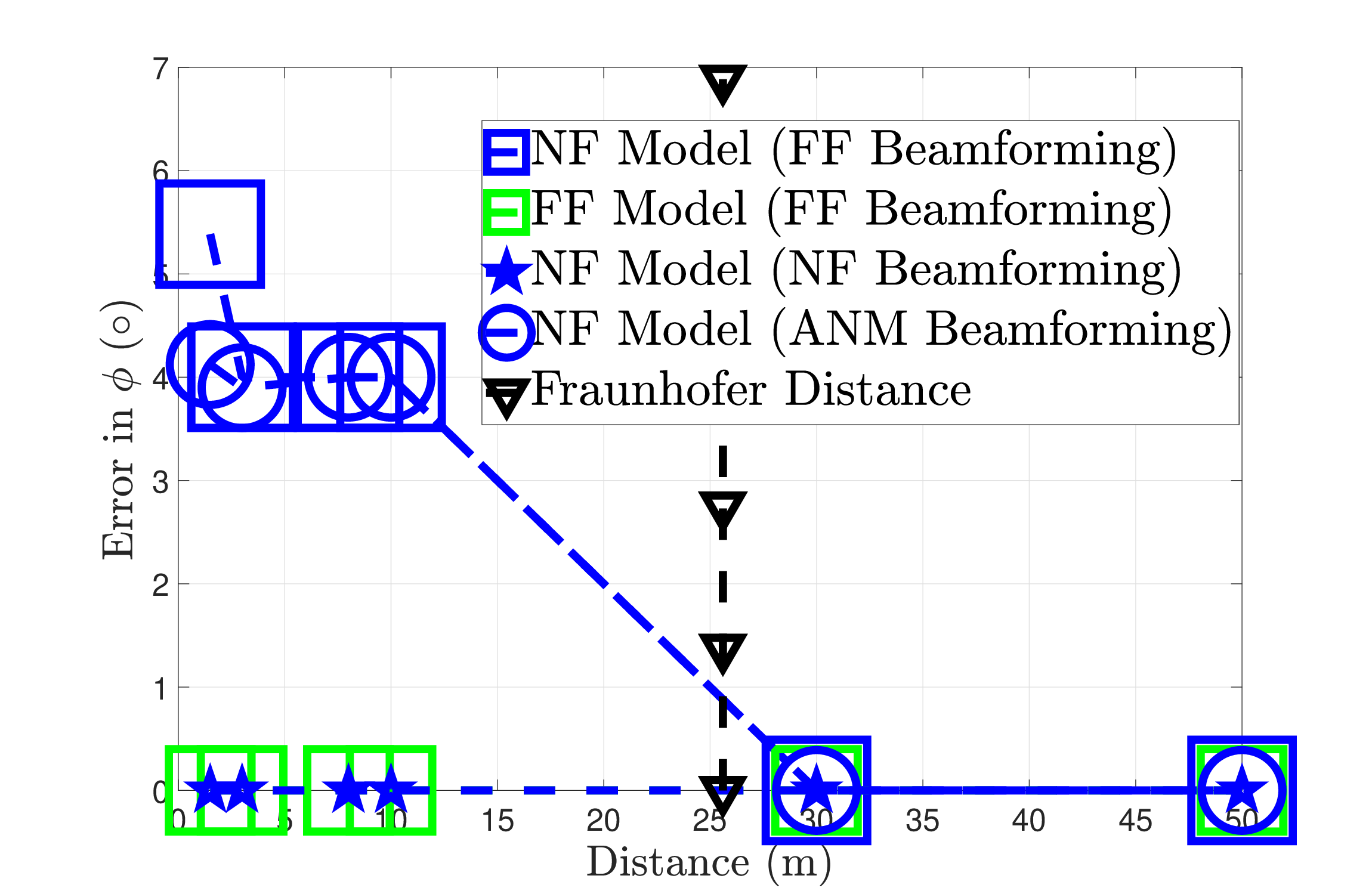}
\label{Figure/NB_2_NS_1280_Nangles_8_fc_1_SNRInd_4_NU__a/Phi_NB_2_NS_1280_Nangles_8_fc_1_SNRInd_4_NU_4.eps}}
\hfil
\subfloat[]{\includegraphics[width=\linewidth]{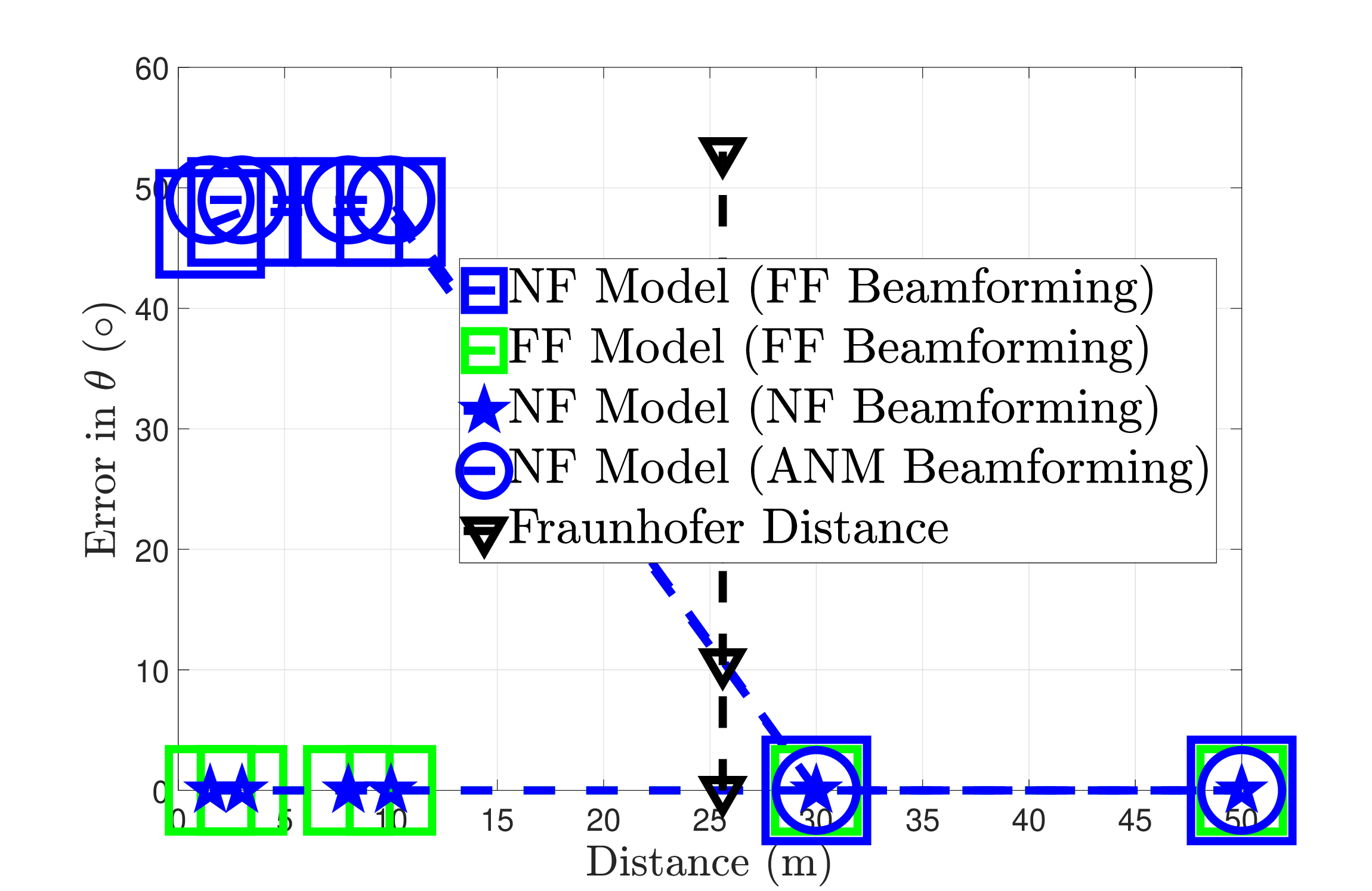}
\label{Figure/NB_2_NS_1280_Nangles_8_fc_1_SNRInd_4_NU__a/Theta_NB_2_NS_1280_Nangles_8_fc_1_SNRInd_4_NU_4.eps}}
\caption{Error in estimating azimuth and elevation angles with $f_c = 3 \times 10^9$, SNR of $30 \text{ dB}$, and $N_U = 256$. This figure highlights the different combinations of the received signal models and the various beamforming vectors.}
\label{Figure/NB_2_NS_1280_Nangles_8_fc_1_SNRInd_4_NU__a/Theta_Phi_NB_2_NS_1280_Nangles_8_fc_1_SNRInd_4_NU4}
\end{figure}

\section{MUSIC with and without Mismatch}
MUSIC relies heavily on the theory of self-adjoint matrices. After collecting the $T$ transmissions received over $N_U$
antennas as $\bm{Y} = [\bm{y}_{1}, \bm{y}_{2}, \cdots, \bm{y}_{T}]$, we write the received signal as
$$
\bm{Y} = \bm{H}\bm{X} + \bm{N}, 
$$
where $\bm{X} = [\bm{x}_{1}, \bm{x}_{2}, \cdots, \bm{x}_{T}]$ and $\bm{N} = [\bm{n}_{1}, \bm{n}_{2}, \cdots, \bm{n}_{T}].$ The covariance matrix of the received signal can be written as
$$
\bm{R}_{\bm{Y}} = {\bm{H}}\bm{R}_{{\bm{X}}}{\bm{H}}^{\mathrm{H}} + N_0\bm{I}_{N_U}.
$$
Here, $\bm{R}_{{\bm{X}}}$ is the covariance matrix of the transmitted signals, and is a diagonal matrix under the assumption of independent transmission from the $N_K$ sources. The $k^{\text{th}}$ column of $\bm{H}$ corresponds to the array response due to the $k^{\text{th}}$ source. The covariance matrix of the received signal can be rewritten as
$$
\bm{R}_{\bm{Y}} = \Tilde{\bm{H}}\Tilde{\bm{R}}_{{\bm{X}}}\Tilde{\bm{H}}^{\mathrm{H}}.
$$
Here, $\Tilde{\bm{H}}$ is the eigenvector matrix and $\Tilde{\bm{R}}_{{\bm{X}}} = \bm{R}_{{\bm{X}}} + N_0\bm{I}_{N_U}$. The first $N_K$ diagonals in $\Tilde{\bm{R}}_{{\bm{X}}}$ is equal to the sum of the eigenvalues of ${\bm{R}}_{{\bm{X}}}$ and $N_0$. The remaining $N_U - N_K$ diagonals are equal to $N_0$. Hence, the last $N_U - N_K$ columns span the noise subspace. Since the signal and noise are selected to be uncorrelated, the following holds
$$
{\bm{H}}^{\mathrm{H}} [\Tilde{\bm{H}}]_{[:,N_K + 1: N_U]} = \bm{0}_{N_K \times (N_U \times N_K) }.
$$
In practice, the covariance matrix, $\bm{R}_{\bm{Y}}$, is unknown, and the correlation matrix is used as an approximation. Hence, the above correlation is not zero. Now, to estimate the DOA (and distance when using the near-field model), we must create beamforming vectors by searching through the possible DOAs and distance values and finding the $K$ DOAs (or $K$ pairs of DOAs and distances when the near-field model is used) that minimize the correlation. These $K$ DOA values are assumed to be the estimate of the $K$ source DOAs (or $K$ pairs of source DOAs and source distances when the near-field model is used). In this paper, there are $3$ possible types of beamforming vectors. Namely, i) $\bm{h}$ when the beamforming matrix is based on the near-field model, $\bm{h}^{\text{ANM}}$ when the beamforming matrix is based on the ANM, and  
$\bm{h}^{FF}$ when the beamforming matrix is based on the far-field model. Clearly, there is a mismatch (and possibly a drop in performance) when $\bm{h}^{\text{ANM}}$ and $\bm{h}^{\text{FF}}$ are used while the destination node is in the near-field region. {\em Another interesting case occurs when  $\bm{h}^{\text{FF}}$ is used, and we incorrectly assume that the destination node is in the far-field region. This case serves to show that we underestimate the DOA estimation error when we incorrectly assume far-field propagation and subsequently use the beamforming vector, $\bm{h}^{\text{FF}}$. This case is also important because it helps to verify if the far-field approximation becomes accurate at far away distances (greater than the Fraunhofer distance) for DOA estimation.}

\begin{figure}[htb!]
\centering
%\subfloat[]{\includegraphics[height=1.8in, width= 1.5in]{Results/ber_ring_ratio_parallel.eps}
\subfloat[]{\includegraphics[width=\linewidth]{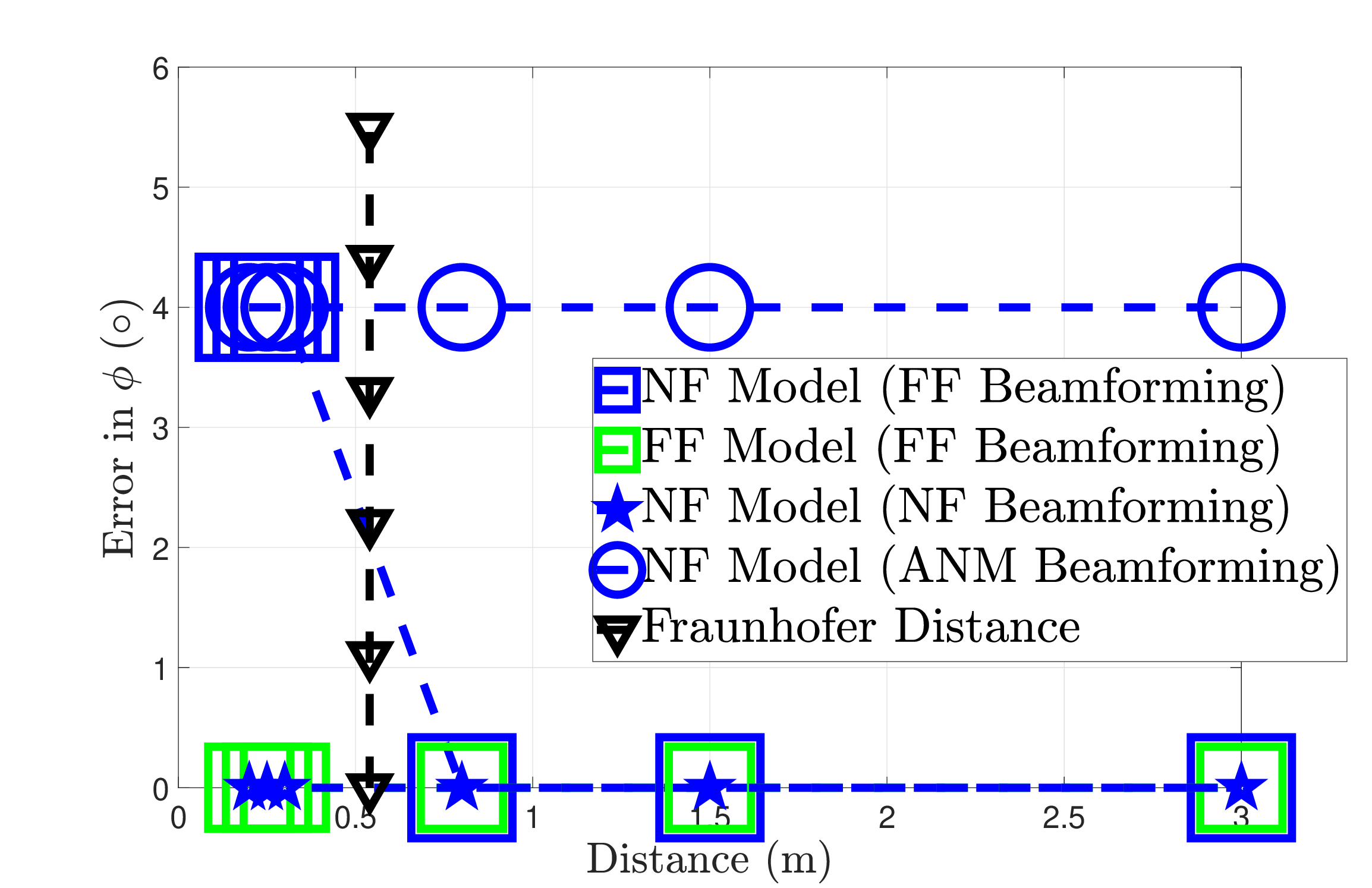}
\label{Figure/NB_2_NS_1280_Nangles_8_fc_4_SNRInd_4_NU__a/Phi_NB_2_NS_1280_Nangles_8_fc_4_SNRInd_4_NU_3.eps}}
\hfil
\subfloat[]{\includegraphics[width=\linewidth]{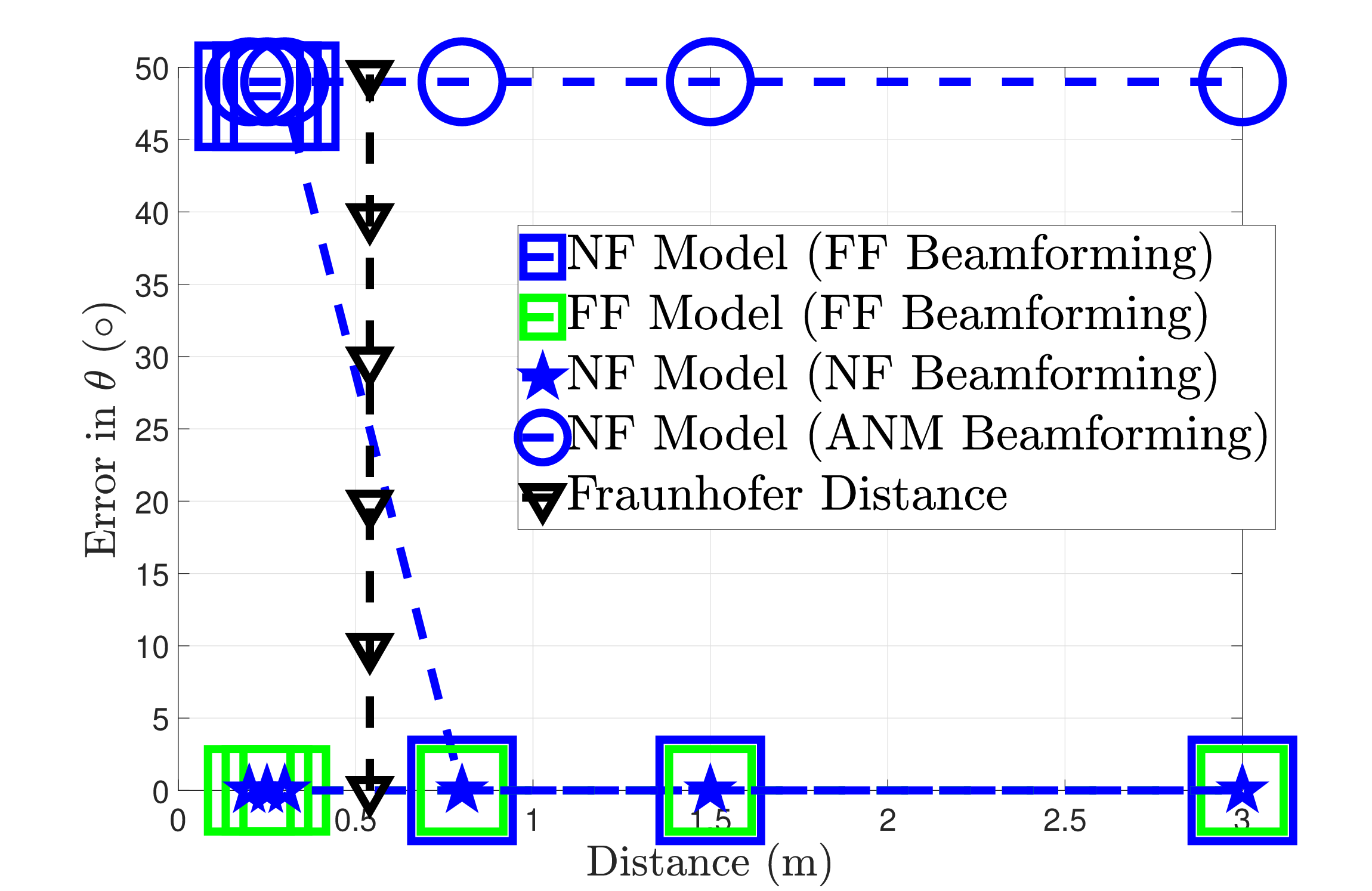}
\label{Figure/NB_2_NS_1280_Nangles_8_fc_4_SNRInd_4_NU__a/Theta_NB_2_NS_1280_Nangles_8_fc_4_SNRInd_4_NU_3.eps}}
\caption{Error in estimating azimuth and elevation angles with $f_c = 80 \times 10^9$, SNR of $30 \text{ dB}$, and $N_U = 144$. This figure highlights the different combinations of the received signal models and the various beamforming vectors.}
\label{Figure/NB_2_NS_1280_Nangles_8_fc_1_SNRInd_4_NU__a/Theta_Phi_NB_2_NS_1280_Nangles_8_fc_4_SNRInd_4_NU3}
\end{figure}

\begin{figure}[htb!]
\centering
%\subfloat[]{\includegraphics[height=1.8in, width= 1.5in]{Results/ber_ring_ratio_parallel.eps}
\subfloat[]{\includegraphics[width=\linewidth]{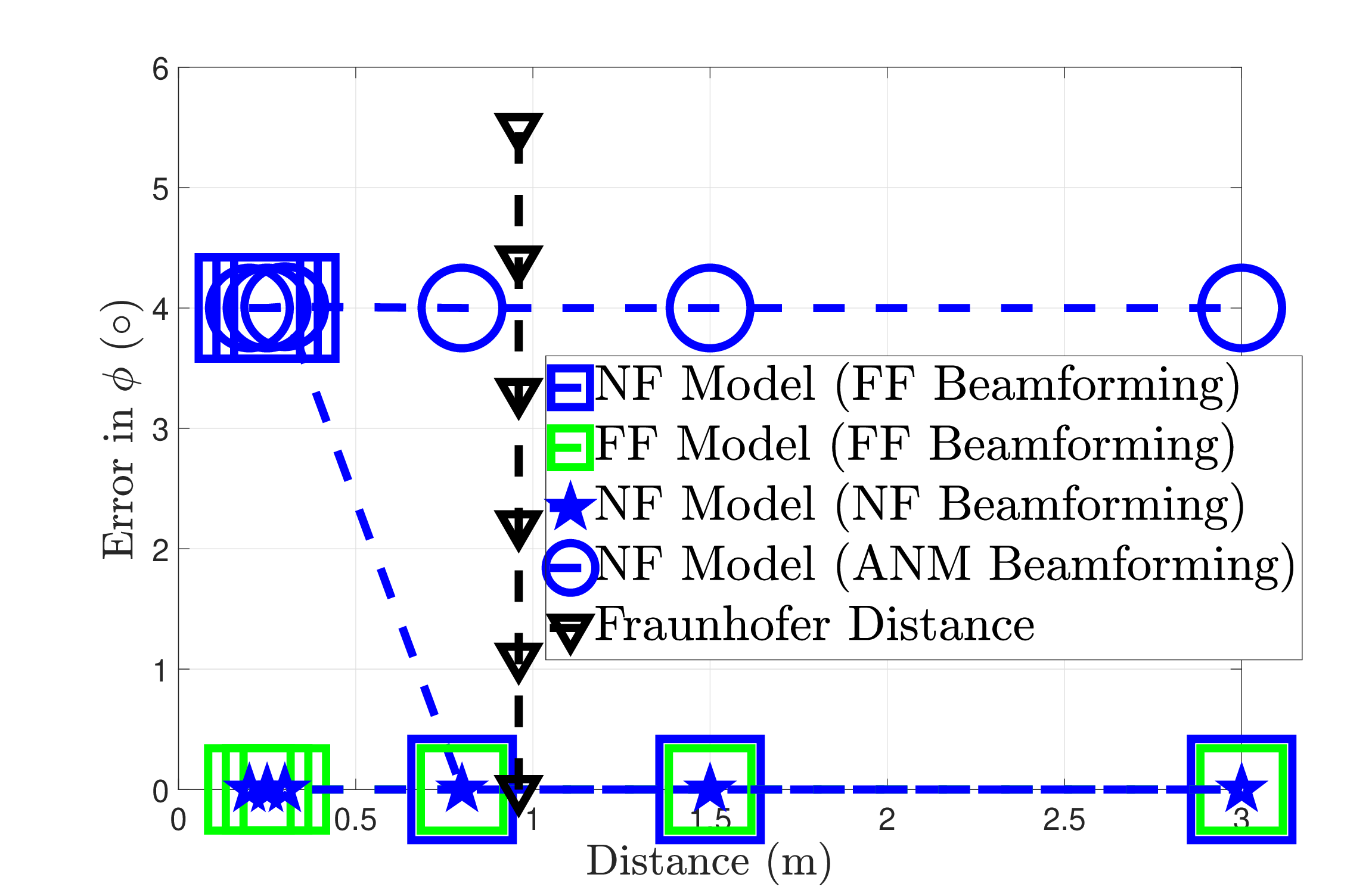}
\label{Figure/NB_2_NS_1280_Nangles_8_fc_4_SNRInd_4_NU__a/Phi_NB_2_NS_1280_Nangles_8_fc_4_SNRInd_4_NU_4.eps}}
\hfil
\subfloat[]{\includegraphics[width=\linewidth]{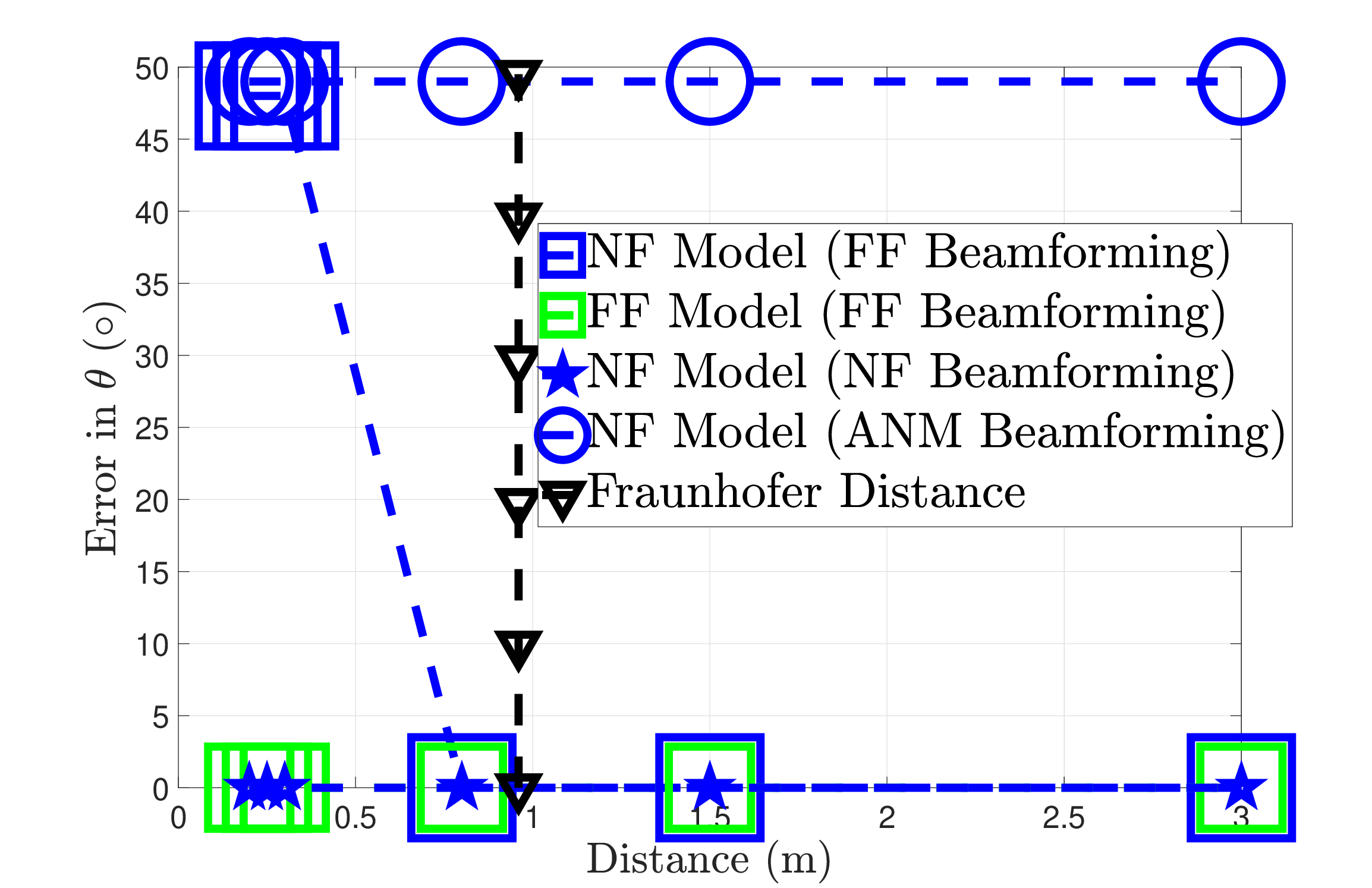}
\label{Figure/NB_2_NS_1280_Nangles_8_fc_4_SNRInd_4_NU__a/Theta_NB_2_NS_1280_Nangles_8_fc_4_SNRInd_4_NU_4.eps}}
\caption{Error in estimating azimuth and elevation angles with $f_c = 80 \times 10^9$, SNR of $30 \text{ dB}$, and $N_U = 256$. This figure highlights the different combinations of the received signal models and the various beamforming vectors.}
\label{Figure/NB_2_NS_1280_Nangles_8_fc_1_SNRInd_4_NU__a/Theta_Phi_NB_2_NS_1280_Nangles_8_fc_4_SNRInd_4_NU4}
\end{figure}

\section{Numerical Results}
In this section, we use simulation results to show the performance of MUSIC under the following scenarios: i) when the beamforming matrix based on the near-field model, $\bm{h}$ is used, while the received signal model is the near-field model, ii) when the beamforming matrix based on the ANM, $\bm{h}^{\text{ANM}}$ is used, while the received signal model is the near-field model, iii)  when the beamforming matrix based on the far-field model,  $\bm{h}^{\text{FF}}$ is used,  while the received signal model is the near-field model, and iv) when the beamforming matrix based on the far-field model  $\bm{h}^{\text{FF}}$ is used, while the received signal model is the far-field model. In our simulations, we consider frequencies of $f_c = 3 \text{ GHz}$ and $f_c = 80 \text{ GHz}$, and the destination location is at $p_U = [6,8,5]^{\mathrm{T}}.$ There are $K = 2$ sources, each located at azimuth angles of $35^{\circ}$ and $39^{\circ}$, and at elevation angles of $63^{\circ}$ and $14^{\circ}$. We consider that the sources are located at the following distances $d_{kU} = [0.2,0.25,0.3,0.8,1.5,3,8,10,30]$ where $k \in \{1,2\}.$ We consider an SNR of $30 \text{ dB}$ and the following number of receive antennas, $N_U \in \{144,256\}.$ In Fig. \ref{Figure/NB_2_NS_1280_Nangles_8_fc_1_SNRInd_4_NU__a/Theta_Phi_NB_2_NS_1280_Nangles_8_fc_1_SNRInd_4_NU3}, the operating wavelength is $0.1 \text{ m}$ and there are $N_U = 144$ receive antennas. Hence, the Fraunhofer distance is $d_f = 14.4 \text{ m}.$ When the near-field beamforming vector $\bm{h}^{}$ is used with the near-field received signal model, the MUSIC algorithm perfectly estimates the DOA in both the azimuth and elevation. This serves as a baseline for other scenarios as it uses the most information (spherical wavefront curvature) and there is no mismatch. However, when the far-field beamforming vector, $\bm{h}^{FF}$ is used with near-field received signal model, there is substantial error in the near-field, but greater than the Fraunhofer distance, the DOA estimation error in both the azimuth and elevation converges to the DOA estimation error when the correct near-field beamforming vector, $\bm{h}^{}$ is used.

%In Fig. \ref{Figure/NB_2_NS_1280_Nangles_8_fc_1_SNRInd_4_NU__a/Theta_Phi_NB_2_NS_1280_Nangles_8_fc_1_SNRInd_4_NU2}, the operating wavelength is $0.1 \text{ m}$ and there are $N_U = 100$ receive antennas. Hence, the Fraunhofer distance is $d_f = 10 \text{ m}.$ In this case, When the near-field beamforming vector, $\bm{h}^{}$, is used with the near-field signal model, the MUSIC algorithm perfectly estimates the DOA in both the azimuth and elevation. This also serves as a baseline for other scenarios. However, when the far-field beamforming vector, $\bm{h}^{FF}$, is used with the near-field signal model, there is a substantial error in the near-field. Still, greater than the Fraunhofer distance, the DOA estimation error in both the azimuth and elevation converges to the DOA estimation error when the correct near-field beamforming vector, $\bm{h}^{}$ is used.

In Fig. \ref{Figure/NB_2_NS_1280_Nangles_8_fc_1_SNRInd_4_NU__a/Theta_Phi_NB_2_NS_1280_Nangles_8_fc_1_SNRInd_4_NU4}, the operating wavelength is still  $0.1 \text{ m}$, but there are $N_U = 256$ receive antennas. Hence, in Fig.  \ref{Figure/NB_2_NS_1280_Nangles_8_fc_1_SNRInd_4_NU__a/Theta_Phi_NB_2_NS_1280_Nangles_8_fc_1_SNRInd_4_NU4}, the Fraunhofer distance is $d_f = 25.6 \text{ m}$. In these scenarios, we still observe similar trends for different combinations of beamforming vectors and received signal models. It is important to notice that in the two figures, using the $\bm{h}^{ANM}$ is not accurate in the near-field, but its performance gets better at far-away distances. {\em Finally, it is crucial to observe that in all four figures, when the near-field beamforming vector, $\bm{h}^{FF}$ is used with the far-field received signal model, the MUSIC algorithm perfectly estimates the DOA in both the azimuth and elevation. This is important to observe because it highlights the fact that assuming far-field propagation underestimates the DOA estimation error. This case is also important because it verifies that the far-field received signal model becomes accurate at far away distances (greater than the Fraunhofer distance) for DOA estimation.}

In Figs. \ref{Figure/NB_2_NS_1280_Nangles_8_fc_1_SNRInd_4_NU__a/Theta_Phi_NB_2_NS_1280_Nangles_8_fc_4_SNRInd_4_NU3} and \ref{Figure/NB_2_NS_1280_Nangles_8_fc_1_SNRInd_4_NU__a/Theta_Phi_NB_2_NS_1280_Nangles_8_fc_4_SNRInd_4_NU4}, the operating wavelength is $3.7 \text{ cm}$, and there are $N_U = 144$, and $N_U = 256$ receive antennas, respectively. The corresponding  Fraunhofer distance for these two figures are $d_f = 0.54 \text{ m}$ and $d_f = 0.96 \text{ m}$, respectively.  In these scenarios, we still observe similar trends for different combinations of beamforming vectors and signal models.

\begin{figure}[htb!]
\centering
%\subfloat[]{\includegraphics[height=1.8in, width= 1.5in]{Results/ber_ring_ratio_parallel.eps}
\subfloat[]{\includegraphics[width=\linewidth]{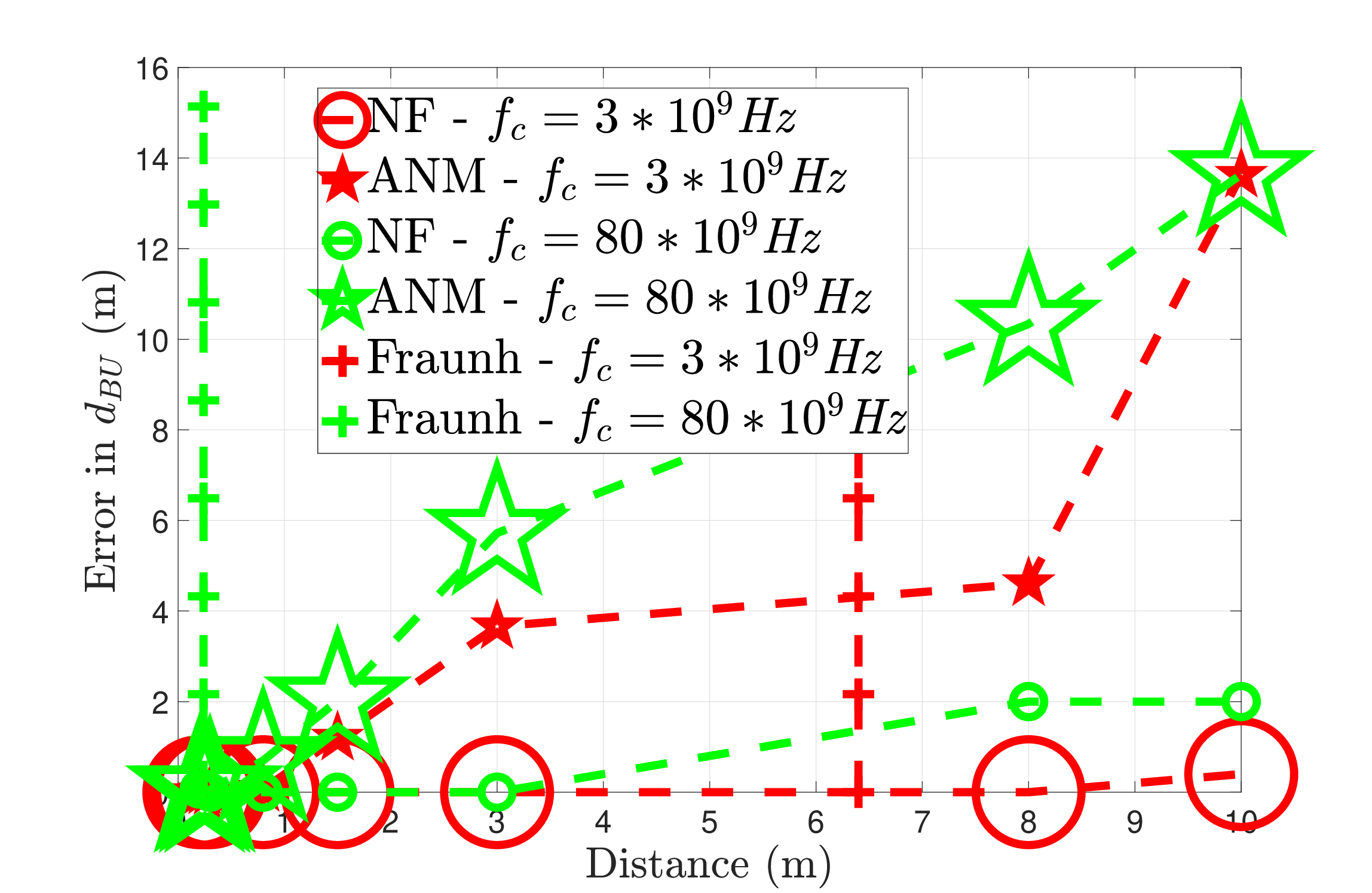}
\label{Figure/Distance/NB_2_NS_1280_Nangles_8_SNRInd_1_NU_1.eps}}
\hfil
\subfloat[]{\includegraphics[width=\linewidth]{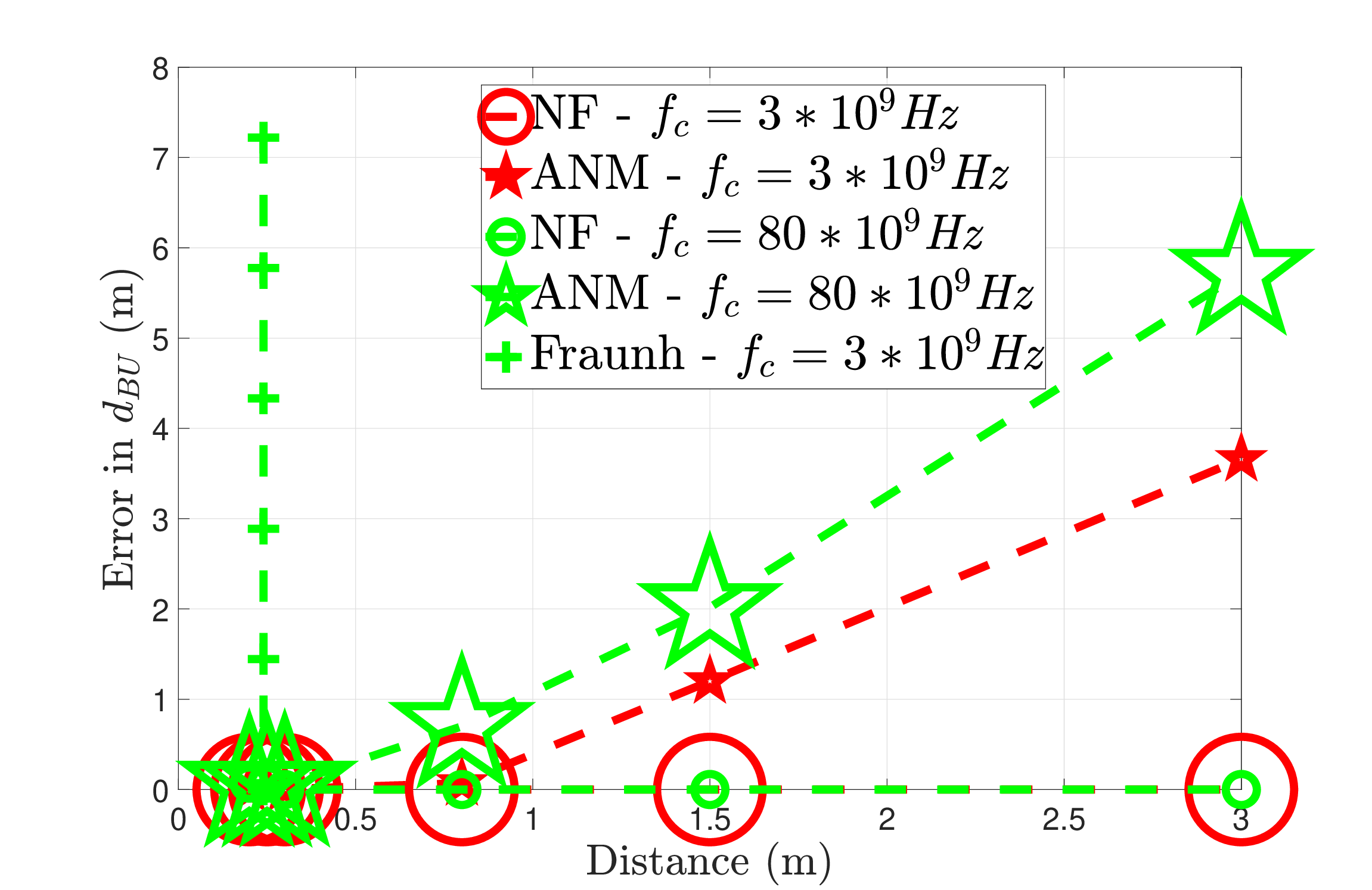}
\label{Figure/Distance/NB_2_NS_1280_Nangles_8_SNRInd_1_NU_1_Short.eps}}
\caption{Error in estimating the relative distance between source and destination with SNR of $10 \text{ dB}$ and $N_U = 64$.}
\label{Figure/Distance/NB_2_NS_1280_Nangles_8_SNRInd_1_NU}
\end{figure}

In Fig. \ref{Figure/Distance/NB_2_NS_1280_Nangles_8_SNRInd_1_NU}, in distances below the Fraunhofer distance, the MUSIC algorithm has almost perfect performance when the near-field beamforming vector $\bm{h}$  is used in estimating the distance. Hence, in the near-field propagation regime, calculating the distance accurately is possible, but in the far-field propagation regime, this accuracy decreases substantially. When the ANM $\bm{h}^{ANM}$ is used, estimating the distance is still possible but much less accurate, even with distances smaller than the Fraunhofer distance. Note that a far-field beamforming vector can not be used to estimate the distance.

\section{Conclusion}
In this paper, we investigated the performance drop when there is a mismatch between the signal model in the MUSIC algorithm and the actual propagation regime. We showed this for DOA and range estimation. More specifically,  we considered the cases when the receiver is in the near-field, but we use: i) the near-field model, ii) the ANM, and iii) the far-field model to design the beamforming matrix in the MUSIC algorithm. Our results indicate that in the near-field, using the ANM and the far-field model to design the beamforming matrix for the MUSIC algorithm causes a reduction in estimation accuracy compared to the case when the correct near-field model is used to design the beamforming matrix. Another result is that the far-field model underestimates the DOA estimation error when we incorrectly assume far-field propagation and subsequently use the far-field beamforming vector. Finally, we showed that the MUSIC algorithm can provide a very accurate estimation of range for distances less than the Fraunhofer distance, but this estimate becomes very inaccurate at distances larger than the Fraunhofer distance.

{
\bibliographystyle{IEEEtran}
\bibliography{refs}
}
\end{document}